\documentclass[
	a4paper, 
	10pt, 
	twoside, 
]{LTJournalArticle}

\usepackage{listings}
\usepackage{float}
\usepackage{comment}

\usepackage{enumitem,amssymb}
\newlist{todolist}{itemize}{2}
\setlist[todolist]{label=$\square$,topsep=0pt}

\usepackage{xcolor}
\usepackage{soul}
\newcommand{\hlc}[2][yellow]{{%
    \colorlet{foo}{#1}%
    \sethlcolor{foo}\hl{#2}}%
}

\definecolor{purple}{HTML}{967BB6}

\usepackage{listings}

\usepackage[most]{tcolorbox}

\newdimen\longline
\longline=\linewidth\advance\longline+4cm

\def\lbl#1{\hbox to 4cm{#1\hfill\strut}}%
\def\labelline#1#2{\lbl{#1}\vbox{\hbox{\raisebox{-1.0\baselineskip}{\TextField[name=#1,width=#2]{\null}}}\kern2pt{\hbox to \longline{\leaders\hbox to 5pt{\hss . \hss}\hfil}}}}

\def\q#1{\hbox to \hsize{\labelline{#1}{\longline}}\vskip0.5ex}

\usepackage{color}

\definecolor{mygray}{rgb}{0.9,0.9,0.9}
\title{SzCORE: A Seizure Community Open-source Research Evaluation framework for the validation of EEG-based automated seizure detection algorithms}

\author{
    Jonathan Dan\textsuperscript{1}\thanks{Corresponding author: \href{mailto:jonathan.dan@epfl.ch}{jonathan.dan@epfl.ch}}\and%
    Una Pale\textsuperscript{1} \and%
    Alireza Amirshahi\textsuperscript{1} \and%
    William Cappelletti\textsuperscript{2} \and
    Thorir Mar Ingolfsson\textsuperscript{3} \and
    Xiaying Wang\textsuperscript{3} \and
    Andrea Cossettini\textsuperscript{3} \and
    Adriano Bernini\textsuperscript{4} \and
    Luca Benini\textsuperscript{3,}\textsuperscript{5} \and
    Sándo Beniczky\textsuperscript{6} \and
    David Atienza\textsuperscript{1} \and
    Philippe Ryvlin\textsuperscript{4}
}

\date{
    \footnotesize\textsuperscript{\textbf{1}}Embedded Systems Laboratory, EPFL, Switzerland\\ %
    \footnotesize\textsuperscript{\textbf{2}}LTS4, EPFL, Switzerland\\ %
    \footnotesize\textsuperscript{\textbf{3}}Integrated Systems Laboratory, ETH Z{\"u}rich, Switzerland\\ %
    \footnotesize\textsuperscript{\textbf{4}}Service of neurology, Centre Hospitalier Universitaire Vaudois, Switzerland\\ %
    \footnotesize\textsuperscript{\textbf{5}}Department of Electrical, Electronic and Information Engineering (DEI), University of Bologna, Italy\\ %
    \footnotesize\textsuperscript{\textbf{6}}Aarhus University Hospital and Danish Epilepsy Centre, Aarhus University, Dianalund, Denmark\\ %
\vspace{1em}
\normalsize\today}

\renewcommand{\maketitlehookd}{%
	\begin{abstract}    
    \begin{minipage}{1.0\linewidth}
        The need for high-quality automated seizure detection algorithms based on electroencephalography (EEG) becomes ever more pressing with the increasing use of ambulatory and long-term EEG monitoring. Heterogeneity in validation methods of these algorithms influences the reported results and makes comprehensive evaluation and comparison challenging. This heterogeneity concerns in particular the choice of datasets, evaluation methodologies, and performance metrics. In this paper, we propose a unified framework designed to establish standardization in the validation of EEG-based seizure detection algorithms. Based on existing guidelines and recommendations, the framework introduces a set of recommendations and standards related to datasets, file formats, EEG data input content, seizure annotation input and output, cross-validation strategies, and performance metrics. We also propose the 10-20 seizure detection benchmark, a machine-learning benchmark based on public datasets converted to a standardized format. This benchmark defines the machine-learning task as well as reporting metrics. We illustrate the use of the benchmark by evaluating a set of existing seizure detection algorithms. The SzCORE (Seizure Community Open-source Research Evaluation) framework and benchmark are made publicly available along with an open-source software library to facilitate research use, while enabling rigorous evaluation of the clinical significance of the algorithms, fostering a collective effort to more optimally detect seizures to improve the lives of people with epilepsy.
    \end{minipage}

    \begin{minipage}{1.0\linewidth}
    \vspace{0.5cm}
        \paragraph{Foreword} This pre-print version of the manuscript intends to collect contributions from the community in order to propose a unified methodology for the validation of seizure detection algorithms in people with epilepsy.
        
        We invite the community to comment on the proposed framework by filling in the following form: \url{https://forms.gle/XfbDaJQi2VooWRN2A}
    \end{minipage}

	\end{abstract}
}

\begin{document}

\maketitle 


\begin{figure*}
	\centering
	\includegraphics[width=\linewidth]{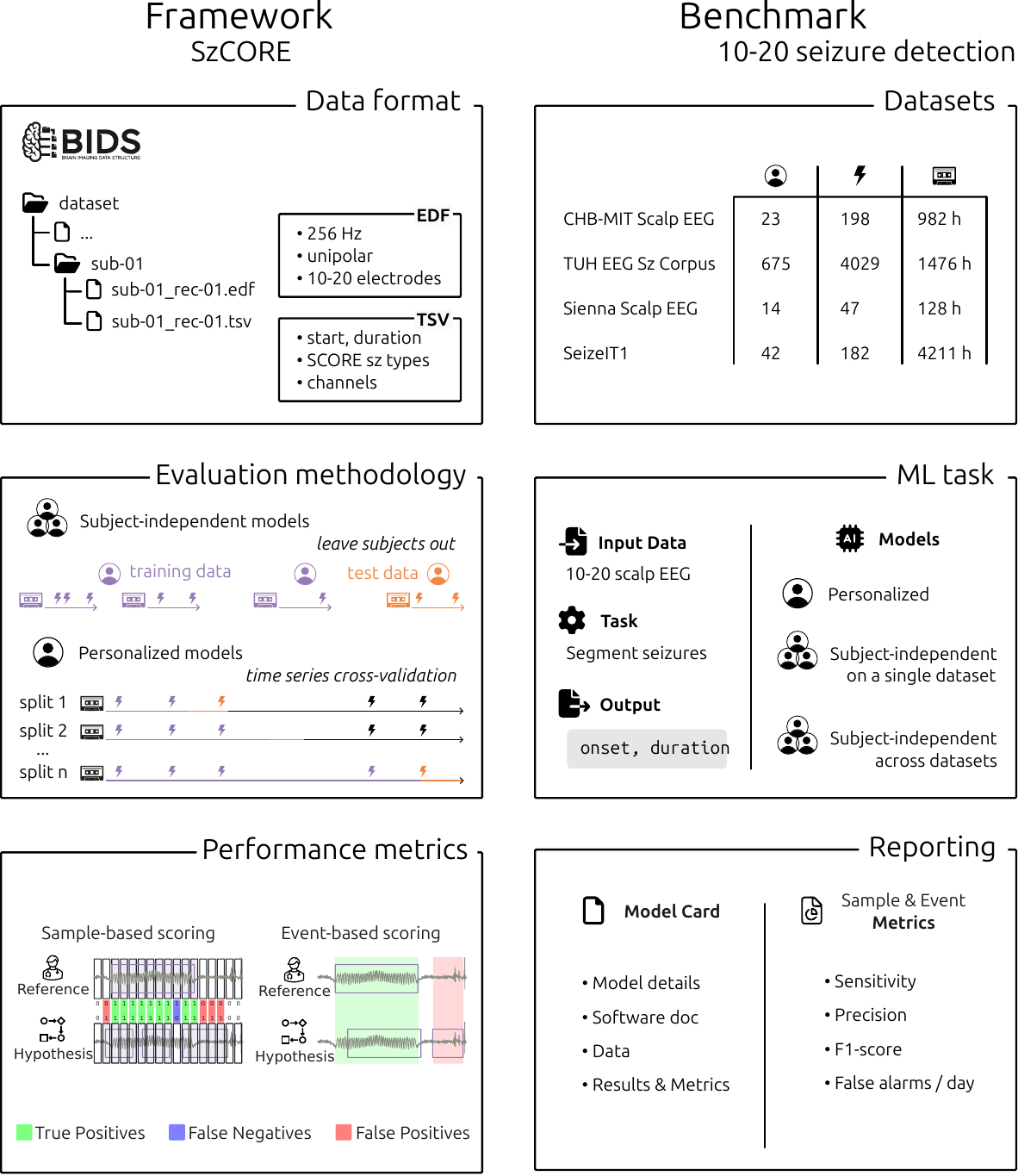}
	\label{fig:graphAbstract}
	\caption*{Graphical abstract}
\end{figure*}


\section{Introduction}
Scalp EEG-based seizure detection algorithms can optimize and facilitate the diagnostic workup performed in people with epilepsy (PWE) to improve patients' care and quality of life~\cite{Baumgartner2018Seizure}. Currently, such algorithms are primarily used during in-hospital long-term video-EEG monitoring (LTM) performed in epilepsy monitoring unit (EMU) over periods of a few days to several weeks. Recordings can be processed on line (i.e. in real time) or off line. Real-time detection helps inform the EMU staff about an ongoing seizure, thus promoting prompt intervention~\cite{Kamitaki2019Yield}, while off-line detection can reduce the physician’s EEG reading workload and help detect subtle seizures.

In the last decade, home-based video-EEG has been gradually developed as an alternative to EMU LTM, which enables the prospect of very long-term ambulatory EEG~\cite{Hasan2021Amulatory}. Home-based video-EEG has similar diagnostic objectives to EMU LTM but can last longer, thanks to lower daily cost and patient and healthcare system burden~\cite{Tatum2021Ambulatory}. It also benefits from automatic seizure detection since it is performed without the permanent supervision of healthcare professionals, and generates large volumes of data.

Ultra long-term ambulatory monitoring has a different scope from LTM and home-based video-EEG recording~\cite{Weisdorf2019Ultra, Macea2023Hospital, Japaridze2022Automated}. It can be used to inform PWE and their caregivers of an ongoing seizure to enable protective interventions, provide physicians with more precise seizure counts than that recalled by PWE and their caregivers to optimize therapy, and document eventual recurrence patterns, which may allow seizure forecasting~\cite{Andrzejak2023Seizure}.

The field of EEG-based seizure detection has benefited from advances in machine learning and the provision of EEG datasets from PWE to train such models. Yet, such datasets with annotated seizures remain rare and often kept private as they must comply with strict legal requirements for personal health data. In contrast, open-source datasets are recognized as catalysts for developing machine-learning algorithms~\cite{Handa2023Eeg}. The machine-learning task can be formulated as a segmentation problem that aims at identifying the start and end of each seizure event. However, current automated scalp EEG-based seizure detection solutions do not meet the level of performance of human experts~\cite{Reus2022Automated}.

A key obstacle hindering progress in the field is the lack of standardized protocols for the training and evaluation of seizure detection algorithms. When developing a novel algorithm, researchers can opt to re-implement selected algorithms for comparison within their own evaluation framework. Such a process is highly time-consuming. Therefore, it is rarely done in practice, resulting in analyses relying on reported metrics that are not necessarily comparable~\cite{Baumgartner2018Seizure}. This issue has been tackled in other research fields by providing a standard machine-learning task definition and benchmark, effectively leading to dramatic improvements in fields such as image classification~\cite{Deng2010ImageNet}, conversational agents~\cite{Rajpurkar2016Squad} or computational models of brain function~\cite{Schrimpf2020Integrative}.

The validation of seizure detection algorithms lacks standardization in EEG datasets, evaluation methodology and performance metrics, as discussed in detail below.

\subsubsection*{EEG datasets} collected for the purpose of individual studies are common in the field~\cite{Baumgartner2018Seizure, Dan2020Computationally,Chatzichristos2022Multimodal}. Such private datasets prohibit direct comparison with studies on other datasets, as algorithm performance is highly data-dependent~\cite{Thuwajit2022EEG}. To date, several datasets have been made publicly available, including Physionet CHB-MIT Scalp EEG Database~\cite{Shoeb2009Application, Goldberger200Physiobank}, TUH EEG Seizure Corpus~\cite{Shah2018The}, Physionet Siena Scalp EEG Database~\cite{Detti2020Siena, Detti2020EEG, Goldberger200Physiobank}, and SeizeIT1~\cite{Chatzichristos2023Seizeit}. Working with multiple datasets is challenging owing to various data formats, e.g. disparities in EEG electrodes, reference electrodes, montage, channel nomenclature, channel sequence, sampling frequencies, and annotation formats. A previous community effort attempted to standardize EEG for computer-based assessment and reporting of EEG, suggesting the SCORE nomenclature, which has been endorsed by the International League Against Epilepsy (ILAE) and International Federation of Clinical Neurophysiology (IFCN)~\cite{Beniczky2017Standardized}. Others have worked on a unified organization of brain imaging files and metadata, suggesting the Brain Imaging Data Structure (BIDS), which is increasingly used in research~\cite{Gorgolewski2016Brain} and which was then extended to organize EEG data~\cite{Pernet2019BIDS}. Recent work has made SCORE machine readable and compatible with BIDS through the HED-SCORE schema specification~\cite{Attia2023Hierarchical}. In subsection~\ref{subsec:datasets} of our framework, \textit{we propose a standard data format for storing EEG and associated seizure annotations that is based on the BIDS-EEG standard and the HED-SCORE nomenclature. The data format provides standardized inputs and outputs for seizure detection algorithms, allowing any seizure detection algorithm to be operated on any thus standardized dataset. Furthermore, this allows visualization and processing of output seizure annotations irrespective of the algorithm that produces them.}

\subsubsection*{Evaluation methodology} has a large influence on reported results. Cross-validation is a statistical method used in machine learning to estimate the performance of an algorithm on an independent dataset~\cite{Refaeilzadeh2009Cross}. To perform cross-validation, the data are split into two sets: a training set and a test set~\footnote{In this paper, we do not cover the notion of a validation set that can be used to determine hyperparameters of a model.}. The performance of an algorithm is reported as the average performance on all test sets after generating multiple models using different splits of training and test data. Many methods exist to split the data, but they do not necessarily meet the requirement of independence between the training and test sets, which could lead to overestimation of the performance of an algorithm. Overestimation of the accuracy of patient-independent models can occur if some of the same subjects are present in both the training and test sets or when datasets are too small~\cite{Shafiezadeh2023Methodological}. Moreover, the chronology of recordings should be respected by only using data in the training set that was acquired prior to the acquisition of the data in the test set for personalized models~\cite{Pale2023Importance}. In subsection~\ref{subsec:cv}, \textit{we propose recommendations for cross-validation of subject-independent and personalized models}.

\subsubsection*{Performance metrics} are critical to estimate the performance of automatic seizure detection. The current use of different metrics makes it difficult to perform comparisons between studies. Reported results use different combinations of general performance metrics, such as sensitivity, specificity, precision, accuracy, area under the receiver operating characteristic curve, f1-score, false-alarm rate, etc. These metrics are computed by comparing ground-truth reference annotations provided by a human expert with hypothesis annotations provided by an algorithm. This comparison allows counting of ``true positives'' (TP; i.e. seizures correctly detected by the algorithm), ``false positives'' (FP; i.e. incorrectly labeled as seizures by the algorithm), and ``false negatives'' (FN; i.e. seizures missed by the algorithm). 

However, TP, FP, and FN can be counted using either \textit{sample-based} scoring or \textit{event-based} scoring, which can result in very different interpretations of the performance metrics. Sample-based scoring computes performance metrics on a sample-by-sample basis and is sometimes referred to as epoch-based scoring~\cite{Shah2021Objective} or window-based scoring. Sample-based scoring is widely adopted by the machine-learning community and it integrates tightly with standard training schemes.  While sample-based scoring captures the fine detail agreement between the reference and hypothesis annotations at the time-scale of labels, it does not provide answers to the clinically relevant questions: \textit{"How many seizures did the patient have?"} or \textit{"How many seizures were missed by the seizure detection algorithm?"} or \textit{"How many false alarms were triggered by the system?"}. Answering these questions requires a scoring method that operates at the granularity level of events (or epileptic seizures), i.e. event-based scoring. This can be computed in many different ways, such as 'Any-overlap' (OVLP) or 'Time-aligned event scoring' (TAES)~\cite{Shah2021Objective}. \textit{In subsection~\ref{subsec:evaluation}, we propose metrics for the evaluation of seizure detection algorithms that are designed to address questions of the clinical community and requirements of the machine-learning community}.

\subsubsection*{}
In summary, the lack of common research practices regarding datasets, cross-validation methodologies and performance metrics when validating seizure detection algorithms is a limiting factor for sound evaluation of algorithms. In this paper, \textit{we propose an open framework for the validation of EEG-based seizure detection algorithms: SzCORE}. This framework is the result of discussions with stakeholders in the field, including PWE, physicians and other healthcare providers, engineers, computer scientists, and other scientists working on the development of seizure detection algorithms. It aims to lift the technical barriers that slow down the development of new algorithms, allowing them to operate on multiple datasets and to be assessed using a fair and objective methodology. Based on the framework, we propose the \textit{10-20 EEG seizure detection benchmark} (Section~\ref{sec:benchmark}) that defines the datasets, tasks and performance evaluation of seizure detection algorithms. Additionally, we provide an open-source code library available on GitHub: \url{https://github.com/esl-epfl/sz-validation-framework}. The library is designed to allow continuous improvement by the community. The framework, benchmark and supporting code library are described on an online platform: \url{https://eslweb.epfl.ch/epilepsybenchmarks}, which also serves as the central hub for a community-built benchmark, where new seizure detection algorithms can be fairly compared.


\section{SzCORE framework}\label{sec:framework}

\subsection{EEG Datasets and data format}\label{subsec:datasets}

\subsubsection{Datasets}
Datasets should include EEG raw signals, recording specifics, seizure annotations, and patient details, e.g. according to BIDS-EEG specifications~\cite{Gorgolewski2016Brain, Pernet2019BIDS}. They should be organized to allow computer systems to process them efficiently. An example of BIDS-EEG data file-structure organization for a dataset of PWE is provided in Appendix~\ref{A:sec:BIDS}.

\subsubsection{EEG data format:}\label{subsec:eegFormat}
To allow algorithms to operate seamlessly on any dataset, we propose standardization of EEG data that is at least consistent with the IFCN and ILAE minimum recording standards that are recommended for EEG~\cite{Peltola2023Routine}. Recordings should be stored in \texttt{.edf} files. They should contain the 19 electrodes of the international 10-20 system in a unipolar common average montage. The recording should be resampled to 256~Hz for storage, and original data should be acquired with a sampling frequency of at least 256~Hz. The channels should be provided in the following order: \texttt{Fp1-Avg, F3-Avg, C3-Avg, P3-Avg, O1-Avg, F7-Avg, T3-Avg, T5-Avg, Fz-Avg, Cz-Avg, Pz-Avg, Fp2-Avg, F4-Avg, C4-Avg, P4-Avg, O2-Avg, F8-Avg, T4-Avg, T6-Avg}. Additional data channels can optionally be provided after these 19 channels; they should not be used to compute the common average.

\subsubsection{Seizure annotation format:}\label{subsec:annotationFormat}
The annotation format should be constructed in a way that it can be used both for original annotations (ground truth) and the output of seizure detection algorithms. The format we propose is a tab-separated values (\texttt{.tsv}) file that is human-readable. It is a text file that uses a tab as a delimiter to separate the different columns of information, with each row representing one event. Each annotation file is associated with a single EEG recording. A detailed description and an example of the information contained in annotation files is provided in appendix~\ref{A:sec:annotation}. These files adhere to the BIDS-EEG guidelines and use the hierarchical ILAE-based classification of seizures defined by HED-SCORE~\cite{Gorgolewski2016Brain, Scheffer2017ILAE, Attia2023Hierarchical}. The seizure nomenclature is presented in Figure~\ref{fig:szType} in Appendix~\ref{A:sec:annotation}.

\subsection{Evaluation methodology}\label{subsec:cv}
To evaluate seizure detection algorithms, a training set is used to determine the parameters of the machine learning algorithm and an independent test set is used to estimate the performance of the algorithm. These sets should be independent to guarantee that results can be generalized to other data. If data are only available from a single setting, the dataset can be split into a training set and a test set. This process is repeated multiple times (i.e. folds) to obtain robust estimates of performance by rotating data between the training set and the test set, i.e. cross-validation~\cite{Refaeilzadeh2009Cross}.

\subsubsection{Personalized models} are trained for a specific patient. These models should successfully detect seizures in unknown recording sessions that took place after the model was initially trained. \textit{To evaluate these models, at each fold, the training set should only include data that was acquired prior to the acquisition of the test set;} this is referred to as time-series-cross-validation (TSCV). 

\begin{figure*} 
	\centering
	\includegraphics[width=\linewidth]{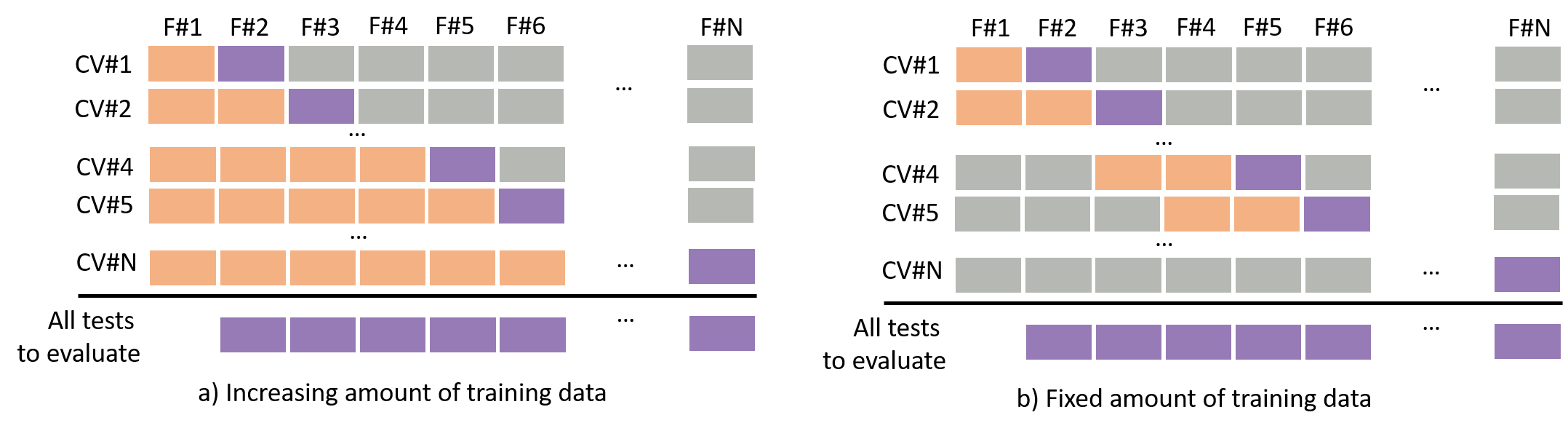}
	\caption{Time series cross-validation for personalized models. Each box represents an epoch of data. \hlc[orange!50]{Orange boxes} are used for training, \hlc[purple!50]{purple boxes} are used for testing. Each row represents a cross-validation fold. The final results are calculated by appending all cross-validation folds (shown in the last row). a) cross-validation scheme with variable amount of data. b) cross-validation scheme with fixed amount of data.}
	\label{fig:persCV}
\end{figure*}

TSCV can be performed in two ways: 
\begin{itemize}
	\item Training data increase as the model is evaluated on future test  folds (variable amount of data, Fig. ~\ref{fig:persCV}a).
	\item Training data keeps a fixed size with past folds removed from the training data as the model is evaluated on future folds (fixed amount of data, Fig. ~\ref{fig:persCV}b).
\end{itemize}

\subsubsection{Subject-independent models} are designed to operate on data from any patient and seizure type. These models should successfully detect seizures in subjects whose data were not used to train the model.

Several methods can be used to validate subject-independent models, provided that independence of subjects between training and test sets is maintained.:
\begin{itemize}
	\item \textit{Leave-one-subject-out (LOO)} is a technique in which many different models are trained~\cite{Hastie2009Model}. Each model is trained using all the data except those from one subject. The data from that subject is used for testing. This allows maximization of the amount of training data provided to the model. Final performance is reported by averaging the testing results of all subjects (each using their subject-independent model). This strategy also allows assessment of the performance of each subject, which can then be compared between different algorithms. However, the technique is not appropriate for large datasets with many subjects, as training models can be computationally expensive and need to be retrained for every subject.
	\item \textit{K-fold cross-validation} uses a similar strategy to LOO~\cite{Hastie2009Model}. The dataset is split into a training and testing subset with a ratio of subjects of $(K-1)/K$ for the training set and $1/K$ for the test set. This split is repeated $K$ times until all subjects are included once in the test set. For each split, a model is trained and performance is reported as an average of each model. This is faster to train and test and, thus, more appropriate for larger datasets as the number of splits is determined by $K$, irrespective of the number of subjects. However, this method uses less data in the training set than LOO, which can lead to sub-optimal models with larger variability in estimated performance. LOO is a special case of K-fold, where $K$ is equal to the number of subjects.
	\item \textit{Fixed training and test sets} with predetermined subjects in each set are appropriate for large datasets (e.g. TUH EEG Sz Corpus). However, it can lead to more variability in estimated performance in small datasets. 
\end{itemize}

While cross-validation allows a fair assessment of algorithms during development, the performance of algorithms for real-world use should be evaluated on large independent datasets which are currently missing in our community.

\subsection{Performance metrics}\label{subsec:evaluation}
To assess the performance of seizure detection algorithms, we propose two complementary scoring methodologies, sample-based and event-based scoring. Both these scoring metrics should be reported when communicating results of algorithms as sampled-based metrics provide a high granularity to machine-learning experts and event-based metrics provide clinically relevant results.

\subsubsection{Sample-based scoring} 
\begin{figure}
	\centering
	\includegraphics[width=\linewidth]{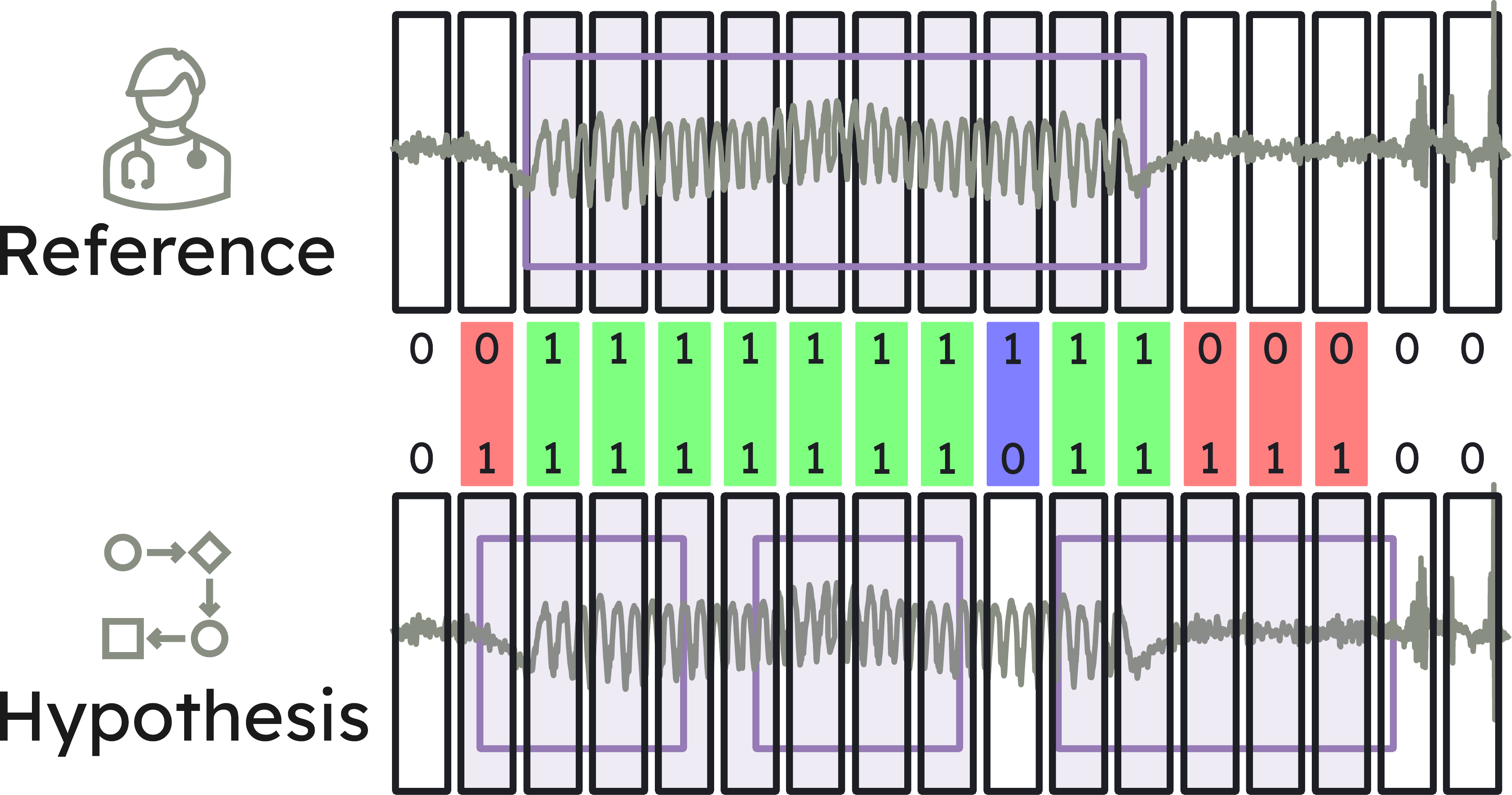}
	\caption{Sample-based scoring compares annotation labels sample by sample. Correct detections \hlc[green!30]{(True Positives)}, false detections \hlc[red!30]{(False Positives)}, missed detections \hlc[blue!30]{(False Negatives)}. Seizure annotations are indicated in purple.}
	\label{fig:sample}
\end{figure}

compares annotation labels, which are provided at a fixed frequency (we propose 1~Hz), sample by sample to detect TP, FP and FN, as shown in figure~\ref{fig:sample}. We propose a frequency of labels of 1~Hz, as it corresponds to the resolution expected by a human annotator. It should be noted, this frequency does not dictate the duration of data windows used to generate machine-learning predictions. These can use an arbitrary duration and overlap as long as they provide predictions at 1~Hz. For annotation labels that overlap only partially with epileptic seizures, we propose to assign a ``seizure'' label to a sample if the overlap exceeds 50\%.

\subsubsection{Event-based scoring} 
\begin{figure}
	\centering
	\includegraphics[width=\linewidth]{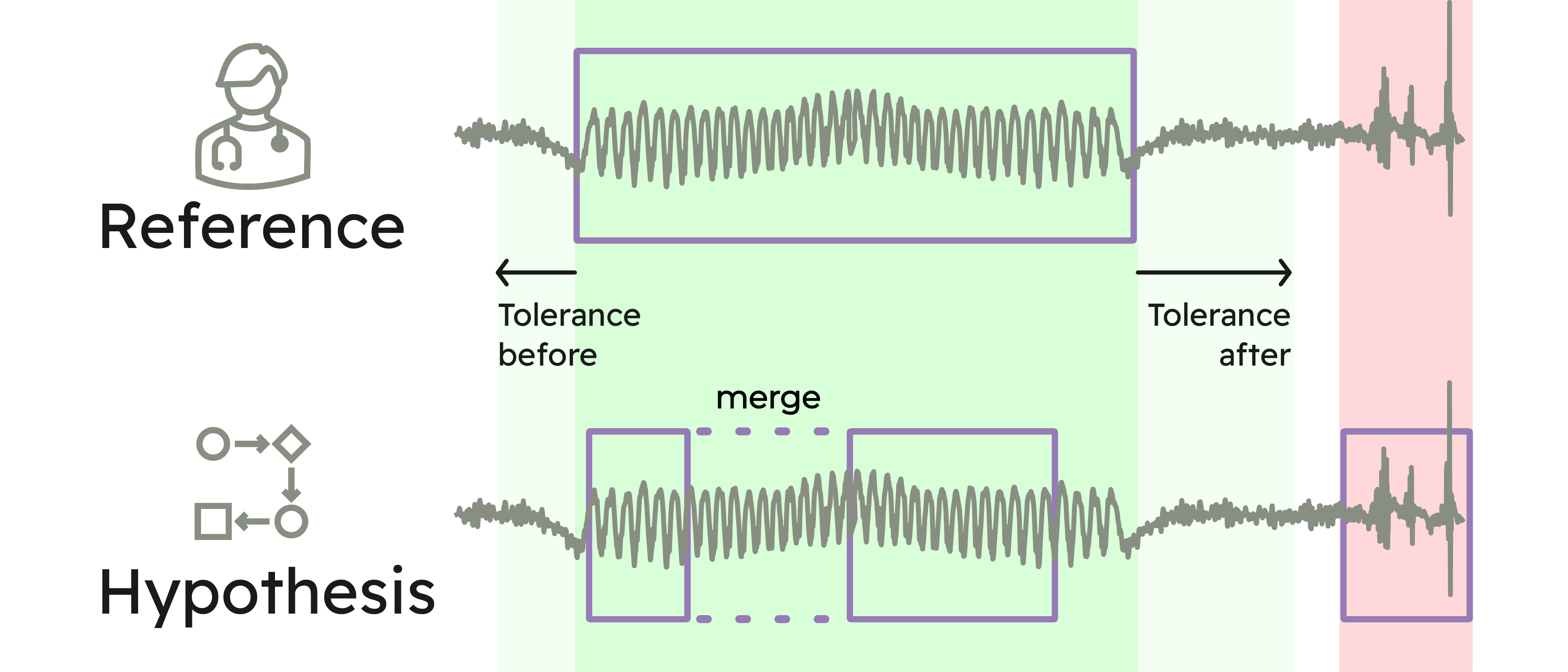}
	\caption{Event-based scoring is based on overlap. It defines a set of rules for event merging, tolerance before and after events, and maximum event duration. Correct detections \hlc[green!30]{(True Positives)}, false detections \hlc[red!30]{(False Positives)}. Seizure annotations are indicated in purple}
	\label{fig:event}
\end{figure}

in which events are seizures, relies on overlap between reference and hypothesis annotations (Figure~\ref{fig:event}). Overlap is considered as correct detection, i.e. TP. If the hypothesis event does not overlap with a reference event, it is counted as FP. 

Accurate annotations of epileptic seizures marking a clear start and end is notoriously difficult. This may be complicated by gradual changes in EEG at the beginning and end of seizures or by other factors, e.g. muscle or movement artifacts. Subtle EEG changes prior to the marked seizure onset or following marked offset are often detected by various algorithms~\cite{Maimaiti2022Overview, Shoeb2011Machine}. Some tolerance is therefore required with regard to the start and stop time of seizure to match annotations between two reviewers (e.g. computer algorithm and human expert). From a practical perspective, many applications of seizure detection algorithms should not be negatively impacted if the algorithm marks seizures slightly earlier or a bit longer than ground-truth annotations. On the contrary, early detection could be beneficial to the patient when the detection algorithm serves as an alarm. 

Another issue concerns seizure duration. As most seizures do not occur in rapid succession, it is reasonable to merge annotations separated by only a few seconds. Finally, because seizures are only exceptionally longer than five minutes (longer events are defined as a status epilepticus~\cite{Trinka2015Definition}) long events are split into multiple events of a maximum of 5 minutes.

These considerations are encoded into the following additional rules and parameters to count seizures:

\begin{description}
	\item[Minimum overlap] Minimum overlap between the reference and hypothesis for a detection. We use any overlap, however short, to enhance sensitivity.
	\item[Pre-ictal tolerance] Tolerance with respect to the start time of an event that would count as a detection. We advise a 30 seconds pre-ictal tolerance.
	\item[Post-ictal tolerance] Tolerance with respect to the end time of an event that would still count as a detection. We advise a 60 seconds post-ictal tolerance.
	\item[Minimum duration between events] Automatically merge events that are separated by less than the given duration. We advise merging events separated by less than 90 seconds which corresponds to the combined pre and post-ictal tolerance.
	\item[Maximum event duration] Split events longer than a given duration into multiple events. We advise splitting events longer than 5 minutes.
\end{description}

\subsubsection{Performance metrics:} Both the sample-based scoring and event-based scoring produce a count of correct detections (TP), missed detections (FN) and wrong detections (FP). These can be used to compute common performance metrics, as defined below. Specifically, sensitivity and precision are of high interest. F1-score is used as a combined measure containing information on both sensitivity and precision.

\begin{description}
	\item[Sensitivity] Percentage of reference seizures detected by the hypothesis. Computed as: $TP / (TP + FN)$
	\item[Precision] Percentage of correct detections over all hypothesis events. Computed as: $TP / (TP + FP)$
	\item[F1-score] Harmonic mean of sensitivity and recall. Computed as: $2 * sensitivity * precision / (sensitivity + precision)$
	\item[False alarms per day] Number of falsely predicted (FP) seizure events, averaged or interpolated to number per day. 
\end{description}

We explicitly \textit{avoid using metrics that rely on a count of TN, such as specificity and accuracy. This is because in the context of event-based scoring, non-seizure events are ill-defined, and in the context of sample-based scoring, non-seizure samples are much more numerous than seizure samples given the rarity of seizures, resulting in extremely high scores for specificity and accuracy, with little clinical relevance.} Errors of the algorithms

\section{Benchmark}\label{sec:benchmark}

The framework described above allows to build a standard by which seizure detection algorithms can be compared. Here, we propose a \textit{10-20 seizure detection benchmark}. That defines:

\begin{itemize}
	\item The data that should be used when evaluating algorithms.
	\item The task and different scenarios that the algorithms should analyze.
	\item The performance metrics and reporting guidelines for these algorithms.
\end{itemize}

\subsection{Benchmark datasets}\label{sec:benchmark_datasets}

Datasets should be publicly and freely available to allow reproducibility testing. Currently, four large public datasets are available~\cite{Wong2023Eeg, Chatzichristos2023Seizeit}, namely Physionet CHB-MIT Scalp EEG Database, TUH EEG Seizure Corpus, Physionet Siena Scalp EEG, and SeizIT1. A summary of the data contained in these datasets is provided in Table~\ref{tab:datasets}.

\begin{table*} 
	\caption{Publicly available scalp EEG datasets of people with epilepsy.}
	\centering
	\begin{tabular}{l | l l l | l l | l l}
		\toprule
		& \multicolumn{3} {c|}{Overview}                         & \multicolumn{2} {c|}{Recordings}  & \multicolumn{2} {c}{Data} \\
		Dataset  & \texttt{\#} subjects & duration & \texttt{\#} seizures & \texttt{\#} files & avg. duration & fs~[Hz]     & \texttt{\#} channels \\
		\midrule
		CHB-MIT  & 23                   & 982~h    & 198                  & 686               & 60~min        & 256         & 22--38               \\
		TUH      & 675                  & 1476~h   & 4029                 & 7377              & 10~min        & [250--1000] & 17--128              \\
		Siena    & 14                   & 128~h    & 47                   & 41                & 150~min       & 512         & 35--45               \\
		SeizeIT1 & 42                   & 4211~h   & 182                  & 458               & 612~min       & 250         & 26                   \\
		\bottomrule
	\end{tabular}
	\label{tab:datasets}
\end{table*}

The currently available public EEG datasets do not all meet the minimum recording requirements of the framework. To use them, the following manipulations are required:
\begin{itemize}
	\item EEG signals are resampled to 256~Hz.
	\item Channels are renamed and rereferenced to 10-20 EEG with a common average reference.
	\item Annotations are converted to BIDS-EEG/HED-SCORE compliant \texttt{.tsv} files.
	\item Data are reorganized according to BIDS-EEG specifications.
	\item Some recordings of the TUH EEG Seizure corpus do not contain all 19 electrodes from the 10-20 system. Missing electrodes are replaced by zero values.
\end{itemize}
An exception is the Physionet CHB-MIT Scalp EEG Database which provides only bipolar channels for which a conversion to the proposed unipolar montage is not possible. This dataset is analyzed with the original bipolar montage.

\subsection{Machine-learning task}\label{sec:benchmark_task}

The machine-learning task can be formulated as a segmentation problem that aims at identifying the start and end of each seizure event. Three test scenarios are proposed for the evaluation of seizure detection algorithms:

\begin{enumerate}
	\item Personalized models
	\item Subject-independent models evaluated on a single dataset
	\item Subject-independent models evaluated across datasets
\end{enumerate}

\subsubsection{Personalized models} require sufficient data per subject in terms of number of seizures ($\geq 3$)\footnote{Three seizures allow at least one seizure for training, validation and test set.} and duration ($\geq 1h30$)\footnote{Two hours correspond to 30min of data around each seizure.} to be effectively trained and evaluated. For this reason, only the following datasets are considered: CHB-MIT, Siena, SeizeIT\footnote{TUH Seizure dataset is excluded as it does not contain enough data (10 minutes on average) per subject and less than three seizures per subject.}. TSCV with a variable amount of data is used. The initial training set includes at least five hours and a minimum of one seizure. Performance is evaluated on the following hour. The process is repeated by successively adding one hour of training data and testing on the next hour until the end of the recording. Performance per subject is calculated for sample and event-based metrics by aggregating all one-hour test sets. The performance of a dataset is computed as the average performance of individual subjects.

\subsubsection{Subject-independent models evaluated on a single dataset} should use LOO or K-fold cross-validation as long as subject-independence is guaranteed. Sample-based metrics aggregate all samples of individual subject. Overall performance is reported as the average of all subjects. Event-based metrics aggregate all events in the same manner. All four datasets can be evaluated. However, for the TUH EEG Seizure Corpus, the fixed split of training, validation and test data provided by the original dataset is used.

\subsubsection{Subject-independent models evaluated across datasets} are trained on a single dataset and tested on the other datasets to verify generalization properties. Sample-based metrics aggregate all samples of individual subject, and then calculate mean performance over all subjects. Event-based metrics aggregate all events in the same manner.

\subsection{Reporting}

The algorithm should report performance for sample-based and event-based scoring including sensitivity, precision, F1-score and false-alarms per day for each individual subject (if possible) and overall average of all subjects. In addition, algorithms should provide enough details to allow result reproducibility, e.g. in a model card including model description, software and environment documentation, data used, evaluation metrics, and results~\cite{Mitchell2019Model}. An example of such a model card is provided in Appendix~\ref{A:sec:checklist}. To help authors document and report results we provide a checklist for reproducible SzCORE algorithms which can be found in Appendix~\ref{A:sec:checklist}.

To test the validity of the framework and as an initial contribution to the benchmark, we ran SzCORE with three algorithms. The performance results of these algorithms are presented in appendix~\ref{A:sec:benchmark}.

\section{Open source library \& benchmark platform}
Along with a description of the framework and benchmark, we provide an open-source code library available on GitHub: \url{https://github.com/esl-epfl/sz-validation-framework}. In its present form, the library provides functionality to perform the following actions.
\begin{itemize}
	\item Convert EEG data from the main public datasets to standardized BIDS-EEG compliant format.
	\item Convert seizure annotations from the main public datasets to standardized HED-SCORE compliant format.
	\item Computing the performance of algorithms using event- or sample-based metrics.
\end{itemize}

The framework, benchmark and supporting code library are described on an online platform: \url{https://eslweb.epfl.ch/epilepsybenchmarks}, which also serves as the central hub for a community-built benchmark of seizure detection algorithms. The platform allows researchers to upload results of a seizure detection algorithm following the framework and benchmark described here. All results are presented in comparative tables and charts. The platform is designed to allow continuous improvement by the community.

\section{Discussion}
In this paper, we present SzCORE, a framework for the validation of EEG-based seizure detection algorithms, and suggest common future research practices, with the aim of allowing fair comparison of performance results and increasing reproducibility of studies. This framework is the result of in-depth discussions with stakeholders from both the medical and computer science communities. 

The present framework defines standards for EEG datasets based on existing guidelines and recommendations. It also defines data formats for EEG and seizure annotations that comply with the BIDS-EEG data organization and HED-SCORE nomenclature. It provides recommendations  and checklist for sound cross-validation of algorithms and defines performance metrics for their evaluation. 

Based on this framework, we propose the \textit{10-20 seizure detection benchmark}. The benchmark defines the datasets, task and performance metrics to evaluate seizure detection algorithms. Additionally, we provide an open-source library to convert data from the public datasets to a standardized data format along with code that implement the performance metrics.

Previous initiatives compared algorithms in the context of contests associated with signal processing congresses (e.g. Neureka IEEE SPMB 2020~\cite{Neureka2020, Chatzichristos2020Epileptic}, ICASSP 2023 seizure detection challenge~\cite{Icassp2023, Irfan2023SeizeFt}). However, evaluation data were not always available after the event, precluding further elaboration or comparison with subsequent algorithms. In contrast, the present benchmark relies on public datasets and it provides a fully transparent evaluation framework, which will hopefully enable continuing progress in the field.

The proposed benchmark could also be compared to existing commercial algorithms, which are still less performant than human experts but have nonetheless already found some use in the clinic~\cite{Reus2022Automated, Koren2021Systematic}. 

The choice of 10-20 scalp-EEG recording content that lies at the core of the present framework is restricted to the minimum recording standards that are recommended for EMU settings~\cite{Peltola2023Routine}. These are, however, not met by some highly promising developments in long-term EEG, particularly ambulatory wearable EEG and subcutaneous EEG, which tend to use a low number of electrodes positioned in non-standard locations~\cite{Macea2023Hospital, Weisdorf2019Ultra}. Whereas our choice appears to exclude such recordings, it can be argued that, whenever possible, recording data with the recommended EMU standards in addition to a novel EEG recording setup guarantees high quality datasets while allowing for the development of specific benchmarks, for example targeting wearable EEG. This was the case for the SeizeIT dataset and ICASSP 2023 seizure detection challenge, which included electrodes positioned behind the ear in addition to standard 10-20 EEG electrodes~\cite{Icassp2023}. In the future, we can expect new guidelines for recording EEG in non-standard locations or different applications that guarantee high-quality datasets. These new recording standards can use the EEG data format defined in this framework such that they integrate seamlessly with the proposed SzCORE evaluation methodology and performance metrics. They will then be used to extend the online platform by setting up new datasets and benchmarks that specifically target those applications.

The presented framework extends previous work that defined seizure scoring~\cite{Shah2021Objective} by complementing sample-based with event-based scoring. The current choice of parameters for these scoring methods is somewhat arbitrary if pragmatic. Ideally, the choice of these parameters should either correspond to a specific use of seizure detection algorithms or be based on known uncertainty. Specific use may require high accuracy, e.g. prompt intervention triggered by seizure alarms. Other uses benefit from high tolerance, e.g. offline review of recordings. In addition, human expert labeling (with is the current gold-standard) shows variation~\cite{Halford2015Inter}, resulting in some uncertainty in labeling the start and end time of seizures~\cite{Maimaiti2022Overview, Shoeb2011Machine}. Our choice in this respect was dictated by the framework, which aims to be generic and fit a wide range of algorithms and application. Some users of the framework might want to adapt some of the parameters to their own use case.

This work effectively addresses some current key issues relating to the validation of seizure detection algorithms~\cite{Shafiezadeh2023Methodological, Pale2023Importance}, including the difficulty in comparing results from different datasets and risks associated with a lack of data independence in cross-validation. The best level of evidence for validation is reached when based on an independent multi-centric dataset with strong generalizability potential. Such a dataset would contain many recordings from different centers from many subjects, including a variety of seizure types, recording equipment, recording protocol, etc. As this may be difficult to obtain, we give recommendation for cross-validation strategies that ensure independence within a single dataset. Future work from the community should aim at collecting a large multi-centric dataset that can be used for the validation of seizure detection algorithms.


\section{Conclusion}
This SzCORE framework and benchmark should foster reproducible, transparent, and efficient research. Crucially, they allow the standardization of the validation of seizure detection algorithms. This will enable direct comparison of reported results that use this benchmark. We also provide well-described performance metrics that are tailored to both the machine-learning and medical communities. The framework, benchmark and accompanying open-source software libraries lower the technical and domain-specific knowledge required for algorithm developers to work on seizure detection algorithms, and test them on multiple datasets. The benchmark will also allow to measure the state of the art of seizure detection algorithms, and guide new research venues.

Moreover, resulting algorithms can serve educational purposes in epilepsy teaching by providing computer-assistant supervision of epileptologists in training worldwide. This is in line with recent recommendations of the Intersectoral Global Action Plan approved by the World Health Organization in 2022, which promotes prioritization of education, training, and improving access to care, including in low- and middle-income countries~\cite{Guekkht2021Road}. 

The benchmark has the potential for further expansion. As more high-quality and ambulatory datasets become available, they can better reflect the range of applications of algorithms. Beyond the detection of epileptic seizures on scalp-EEG, the development of the benchmark can address other EEG features and other physiological signals. 

In order to encourage the adoption of the framework, we have set up a community online platform to describe it and collect results of algorithms that use it \url{https://eslweb.epfl.ch/epilepsybenchmarks}. We welcome any suggestions for new datasets, new tasks, or improvements to the methodology or content.

\section*{Acknowledgements}
The Pedesite consortium participated in this study through critical feedback on the proposed methodology. In particular the following individuals (some in the author list) were involved :  Alireza Amirashi\textsuperscript{1}, David Atienza\textsuperscript{1}, Jonathan Dan\textsuperscript{1}, Jose Miranda\textsuperscript{1}, Una Pale\textsuperscript{1}, Amirhossein Shahbaziniae\textsuperscript{1}, William Cappelletti\textsuperscript{2}, Abdellah Rahmani\textsuperscript{2}, Adriano Bernini\textsuperscript{3}, Alexandre Pfister\textsuperscript{3}, Philippe Ryvlin\textsuperscript{3}, Antoine Spahr\textsuperscript{3}, Simone Benatti\textsuperscript{4, 5}, Luca Benini\textsuperscript{4, 5}, Andrea Cossettini\textsuperscript{4}, Thorir Mar Ingolfsson\textsuperscript{4}, Xiaying Wang\textsuperscript{4}.

\begin{enumerate}
	\item Embedded Systems Laboratory, EPFL, Switzerland
	\item LTS4, EPFL, Switzerland
	\item Service of neurology, Centre Hospitalier Universitaire Vaudois, Switzerland
	\item Integrated Systems Laboratory, ETH Z{\"u}rich, Switzerland
	\item Department of Electrical, Electronic and Information Engineering (DEI), University of Bologna, Italy
\end{enumerate}

In addition we would like to thank the many international collaborators who participated in discussions that helped build this work. In particular the participants of the Fourth International Congress on Mobile Health and Digital Technology in Epilepsy (2023); Christos Chatzichristos, Lauren Swinnen, Jaiver Macea and Nick Seeuws from KU Leuven (Belgium); Bernard Dan, Karine Pelc from ULB (Belgium).

\section*{Author contributions}
\begin{itemize}
    \item Jonathan Dan: Conceptualization, Methodology, Software, Validation, Data Curation, Writing, Visualization, Project administration
    \item Una Pale: Conceptualization, Methodology, Software, Investigation, Writing, Visualization
    \item Alireza Amirshahi: Methodology, Investigation, Writing – Original Draft 
    \item William Cappelletti: Methodology
    \item Thorir Mar Ingolfsson: Methodology, Investigation, Writing – Original Draft 
    \item Xiaying Wang: Writing – Review \& Editing
    \item Andrea Cossettini: Writing – Review \& Editing, Supervision
    \item Adriano Bernini: Methodology
    \item Luca Benini: Writing – Review \& Editing, Supervision, Funding acquisition
    \item Sándor Beniczky: Writing – Review \& Editing, Supervision
    \item David Atienza: Writing – Review \& Editing, Supervision, Funding acquisition
    \item Philippe Ryvlin: Methodology, Writing – Review \& Editing, Supervision, Funding acquisition
\end{itemize}


\printbibliography 


\onecolumn

\appendix


\section{Data format}

\subsection{BIDS-EEG compliant dataset}\label{A:sec:BIDS}

Here we present the file structure organization of the Physionet CHB-MIT Scalp EEG Database converted to BIDS-EEG~\cite{Gorgolewski2016Brain, Pernet2019BIDS}. Annotations from seizure detection algorithms are placed in the \texttt{szDetection} derivatives folder that can be distributed with or without the original dataset. The CHB-MIT dataset converted to BIDS-EEG is made available on zenodo: \url{https://zenodo.org/records/10259996}.

\begin{verbatim}
BIDS_CHB-MIT/
├── README
├── dataset_description.json
├── events.json
├── participants.json
├── participants.tsv
├── sub-01/
│   ├── ses-01/
│   │   └── eeg/
│   │       ├── sub-01_ses-01_task-szMonitoring_run-00_eeg.edf
│   │       ├── sub-01_ses-01_task-szMonitoring_run-00_eeg.json
│   │       ├── sub-01_ses-01_task-szMonitoring_run-00_events.tsv
│   │       ├── ...
│   ├── ...
├── ...
├── szDetection/
│   ├── sub-01/
│   │   └── ses-01/
│   │       ├── sub-01_ses-01_task-szMonitoring_run-00_events.tsv
\end{verbatim}

\subsection{Annotation format}\label{A:sec:annotation}

\begin{figure}
	\centering
	\includegraphics[width=0.7\linewidth]{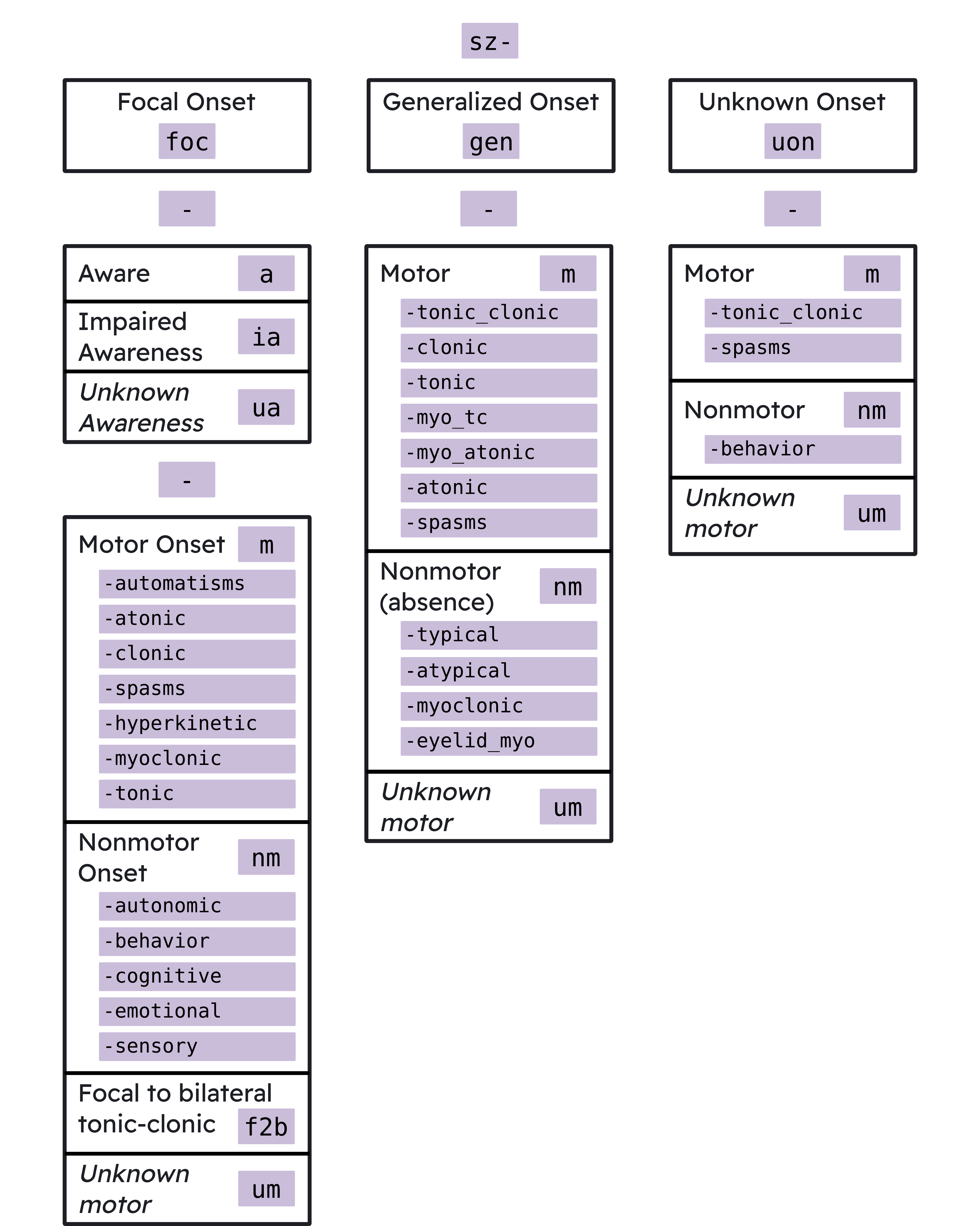}
	\caption{ILAE 2017 Classification of seizure types (expanded version)~\cite{Scheffer2017ILAE} \hlc[purple!50]{Items in purple} are used as short codes to describe an event. As an example a generalized tonic-clonic seizure would be given the code: \texttt{sz-gen-m-tonic\_clonic}.}
	\label{fig:szType}
\end{figure}

The annotation format is a tab-separated values (\texttt{tsv}) file. It is HED-SCORE compliant. It contains the following information:

\begin{description}
	\item[onset] represents the start time of the event from the beginning of the recording, in seconds.
	\item[duration] represents the duration of the event, in seconds.
	\item[event] indicates the type of the event. The event field is primarily used to describe the seizure type. Seizure events begin with the value \texttt{sz}. They can optionally contain more detailed seizure types, as shown in Figure~\ref{fig:szType}. Recordings with no seizures use the string \texttt{bckg} with the event duration equal to the recording duration.
	\item[confidence] represents confidence in the event label. Values are in the range [0--1] [no confidence -- fully confident]. This field is intended for the confidence of the output prediction of machine learning algorithms. It is optional, if it is not provided value should be \texttt{n/a}.
	\item[channels] represents channels to which the event label applies. If the event applies to all channels, it is marked with the value \texttt{all}. Channels are listed with coma-separated values. It is optional, if it is not provided value should be \texttt{n/a}.
	\item[dateTime] start date time of the recording file. The date time is specified in the \texttt{POSIX} format \texttt{\%Y-\%m-\%d \%H:\%M:\%S} (e.g., 2023-07-24 13:58:32). The start time of a recording file is often specified in the metadata of the \texttt{edf}.
	\item[recordingDuration] refers to the total duration of the recording file in seconds.
\end{description}

\begin{lstlisting}[title= An annotation file that contains three seizures.]
 onset	 duration	eventType	confidence	channels	dateTime	         recordingDuration
 296.0	 40.0    	sz      	n/a     	n/a     	2016-11-06 13:43:04	 3600.00
 453.0	 12.0    	sz      	n/a     	n/a     	2016-11-06 13:43:04	 3600.00
 895.0	 21.0    	sz      	n/a     	n/a     	2016-11-06 13:43:04	 3600.00
\end{lstlisting}

We propose to adopt the ILAE classification of seizure types to describe seizure types~\cite{Scheffer2017ILAE} stored in the event field. The classification is hierarchical, depending on available clinical information. At the top level, the seizure type is unspecified (\texttt{sz}). The second level describes the seizure onset zone (focal: \texttt{sz-foc}, generalized \texttt{sz-gen} or unknown \texttt{sz-uon}). Further levels describe the awareness, motor components and seizure symptomology. The full list of standardized seizure types is presented in Figure~\ref{fig:szType}. They are linked to the hierarchy defined by HED-SCORE. The mapping to HED tags is provided in the BIDS-EEG converter library.

\section{Benchmark}\label{A:sec:benchmark}

Here we describe three algorithms that implement the 10-20 seizure detection benchmark.

\subsection{Algorithms}

\subsubsection{Random forest with Approximate zero-cross features} 
This random forest is lightweight model that performs relatively well for epileptic seizure detection. It has been extensively used for EEG-based seizure classification~\cite{Baldassano2017Crowdsourcing,  Sopic2018EGlass, Siddiqui2020Review}. A recent paper by Zanetti et al.~\cite{Zanetti2022Approximate} demonstrated that six approximate zero-crossing (AZC) features can outperform a set of classical literature features (CLF) on two publicly available datasets: CHB-MIT (scalp EEG) and SWEC-ETHZ (intracranial EEG)~\cite{Burrello2019Laelaps}. The hyperparameters are an ensemble of 100 decision trees built with the Gini split criterion. 

\subsubsection{Transformer} 
In this model, a short-time Fourier transform~(STFT) is applied to 12-second windows of EEG. The STFT is computed on one-second segments, 50 samples of overlap, and a frequency resolution of 2 Hz. These parameter choices are extracted from the recommendations in \cite{ma2023tsd}.
The model used for this task is a 4-layer VisionTransformer-based model~\cite{dosovitskiy2020image}, which is modified for epileptic seizure detection by~\cite{ma2023tsd}. The STFT extracted from the EEG input signal is considered as an input image to this 4-layer transformer encoder. The decoder is implemented as a fully connected layer, reducing the dimensions to match the number of classes.

\subsubsection{XGBoost} 
In their seminal work, Ingolfsson et al.~\cite{ingolfsson2021towards} demonstrated the efficacy of Discrete Wavelet Transform (DWT) attributes as robust discriminators for seizure events when integrated into classical machine learning architectures such as Random Forests and Decision Trees. Building upon this foundation, the algorithms enriches this paradigm by incorporating Gradient Boosted Trees — specifically, the advanced eXtreme Gradient Boosting (XGBoost) framework. The approach exploits the discriminative power of DWT features in synergy with XGBoost. In addition,  Approximate zero-cross features explained in the section above are provided to the algorithm. The window size is set to 1 second.

\subsection{Results}\label{sec:results } 

\begin{table}
	\caption{Performance metrics of subject-independent seizure detection algorithms trained on a single dataset and evaluated on the same dataset using cross-validation.}
	\centering
	\begin{tabular}{l|c|cccc|cccc}
		\toprule
		Model & Metrics & \multicolumn{4}{c|}{Event-based} & \multicolumn{4}{c}{Sample-based} \\
		            &             & CHB-MIT & TUH  & Siena & SeizeIT & CHB-MIT & TUH  & Siena & SeizeIT \\
		\midrule
		RF          & F1-score    & 53.9    &      & 59.3  &         & 21.7    &      & 19.8  &         \\
		            & Sensitivity & 37.0    &      & 32.9  &         & 10.8    &      & 6.8   &         \\
		            & Precision   & 64.5    &      & 62.5  &         & 71.9    &      & 69.8  &         \\
		            & FP/day     & 1.66    &      & 1.7   &         & -       &      & -     &         \\
		\midrule
		Transformer & F1-score    &         & 63.2 & 22.8  &         &         & 61.6 & 17.3  &         \\
		            & Sensitivity &         & 76.5 & 56.9  &         &         & 60.7 & 23.3  &         \\
		            & Precision   &         & 53.9 & 18.7  &         &         & 62.4 & 69.9  &         \\
		            & FP/day     &         & 40.6 & 34.4  &         &         & -    & 21.1  &         \\
		\midrule
		XGBoost     & F1-score    & 66.0    &      &       &         & 52.0    &      &       &         \\
		            & Sensitivity & 67.1    &      &       &         & 48.14   &      &       &         \\
		            & Precision   & 75.3    &      &       &         & 52.19   &      &       &         \\
		            & FP/day     & 2.09    &      &       &         & -       &      &       &         \\
		\bottomrule
	\end{tabular}
\end{table}

\begin{table}
	\caption{Performance metrics of subject-independent seizure detection algorithms trained on a dataset and evaluated on independent datasets.}
	\centering
	\begin{tabular}{l|c|c|cccc|cccc}
		\toprule
		Model & Metrics & Training & \multicolumn{4}{c|}{Event-based} & \multicolumn{4}{c}{Sample-based} \\
		            &             & data & CHB-MIT & TUH     & Siena & SeizeIT & CHB-MIT & TUH     & Siena & SeizeIT \\
		\midrule
		RF          & F1-score    & Siena &21.2     &         &  $\vert$     &         &11.1     &         &       &         \\
		            & Sensitivity & Siena &22.2     &         &  $\vert$     &         &7.1      &         &       &         \\
		            & Precision   & Siena &20.4     &         &  $\vert$     &         &20.9     &         &       &         \\
		            & FP/day     & Siena &2.23     &         &  $\vert$     &         &-        &         &       &         \\
		\midrule
  		RF          & F1-score    & CHB-MIT & $\vert$     &         &  41.8  &         &$\vert$     &         &  13.9   &         \\
		            & Sensitivity & CHB-MIT & $\vert$     &         &  48.9  &         &$\vert$     &         &  8.5    &         \\
		            & Precision   & CHB-MIT & $\vert$     &         &  52.5  &         &$\vert$     &         &  64.9   &         \\
		            & FP/day     & CHB-MIT & $\vert$     &         &  9.4   &         &$\vert$     &         &  -      &         \\
		\midrule
		Transformer & F1-score    & TUH  &         & $\vert$ & 32.2  &         &         & $\vert$ & 26.8  &         \\
		            & Sensitivity & TUH  &         & $\vert$ & 54.4  &         &         & $\vert$ & 28.9  &         \\
		            & Precision   & TUH  &         & $\vert$ & 31.3  &         &         & $\vert$ & 35.9  &         \\
		            & FP/day     & TUH  &         & $\vert$ & 21.9  &         &         & $\vert$ & -     &         \\
		\midrule
		XGBoost     & F1-score    &      &         &         &       &         &         &         &       &         \\
		            & Sensitivity &      &         &         &       &         &         &         &       &         \\
		            & Precision   &      &         &         &       &         &         &         &       &         \\
		            & FP/day     &      &         &         &       &         &         &         &       &         \\
		\bottomrule
	\end{tabular}
\end{table}

\begin{table}
	\caption{Performance metrics of subject-specific models}
	\centering
	\begin{tabular}{l|c|cccc|cccc}
		\toprule
		Model & Metrics & \multicolumn{4}{c|}{Event-based} & \multicolumn{4}{c}{Sample-based} \\
		            &             & CHB-MIT & TUH* & Siena & SeizeIT & CHB-MIT & TUH* & Siena & SeizeIT \\
		\midrule
		RF          & F1-score    &         72.7& -    &       58.8&         &         51.7& -    &       36.0&         \\
		            & Sensitivity &         74.0& -    &       62.8&         &         40.0& -    &       26.8&         \\
		            & Precision   &         77.5& -    &       55.6&         &         82.7& -    &       65.2&         \\
		            & FP/day     &         1.58& -    &       4.64&         &         -& -    &       -&         \\
		\midrule
		Transformer & F1-score    &         & -    &       &         &         & -    &       &         \\
		            & Sensitivity &         & -    &       &         &         & -    &       &         \\
		            & Precision   &         & -    &       &         &         & -    &       &         \\
		            & FP/day     &         & -    &       &         &         & -    &       &         \\
		\midrule
		XGBoost     & F1-score    & 79.8    & -    &       &         & 60.5    & -    &       &         \\
		            & Sensitivity & 87.36   & -    &       &         & 54.63   & -    &       &         \\
		            & Precision   & 80.14   & -    &       &         & 77.75   & -    &       &         \\
		            & FP/day     & 0.99    & -    &       &         & -       & -    &       &         \\
		\bottomrule
	\end{tabular}
	{\small{*TUH is not considered due to insufficient data per subject}}
\end{table}

\clearpage

\section{Model Card \& SzCORE Reproducibility Checklist}\label{A:sec:checklist}

In table~\ref{tbl:modelCard}, we provide a template model card for reporting results of seizure detection algorithms. The model card contains contact details, model details and a standardized presentation of performance results. It is provided as an editable PDF, a \LaTeX\ template and a Microsoft Word template. All of which can be downloaded here : \url{https://eslweb.epfl.ch/epilepsybenchmarks/model-card/}.

In table~\ref{tbl:checklist}, we provide a checklist for authors that report results based on the SzCORE framework. The checklist is based on \textit{The Machine Learning Reproducibility Checklist}~\cite{Pineau2021Improving}.

\begin{table}
\caption{Model Card for reporting results of algorithms validated using SzCORE.}
\label{tbl:modelCard}
\begin{Form}
\begin{tcolorbox}[colback=white,colframe=black,colbacktitle=black!10!white, coltitle=black,title={Model name: \raisebox{-0.25\baselineskip}{\TextField[name=title, width=12em]{}}}]
	\subsection*{Contact Details}

		\q{Developers}
		\q{Institution}
		\q{Contact email}
			
	\tcbline
			  
	\subsection*{Model Details}
		\q{Link to source code}
		\q{Citation details}
		\q{Model description}
		\q{ }
			
	\tcbline
			
	\subsection*{Results}
 		\begin{table}[H]
			\centering
			\begin{tabular}{l|r|cccc|cccc}
				\multicolumn{10}{c}{\textit{Performance of a subject-specific model}} \\
				\midrule
				\multicolumn{2}{r|}{Metrics} & \multicolumn{4}{c|}{Event-based} & \multicolumn{4}{c}{Sample-based} \\
				\multicolumn{2}{l|}{} & CHB-MIT & TUH & Siena & SeizeIT & CHB-MIT  & TUH & Siena & SeizeIT \\
				\midrule
				\multicolumn{2}{r|}{F1-score}        & \TextField[name = ps.e.f1.chb,width=2.5em]{}  & -  & \TextField[name = ps.e.f1.siena,width=2.5em]{} & \TextField[name = ps.e.f1.sz,width=2.5em]{} & \TextField[name = ps.s.f1.chb,width=2.5em]{}  & -  & \TextField[name = ps.s.f1.siena,width=2.5em]{} & \TextField[name = ps.s.f1.sz,width=2.5em]{} \\
				\multicolumn{2}{r|}{Sensitivity}     & \TextField[name = ps.e.s.chb,width=2.5em]{}   & -  & \TextField[name = ps.e.s.siena,width=2.5em]{}  & \TextField[name = ps.e.s.sz,width=2.5em]{}  & \TextField[name = ps.s.s.chb,width=2.5em]{}   & -  & \TextField[name = ps.s.s.siena,width=2.5em]{}  & \TextField[name = ps.s.s.sz,width=2.5em]{}  \\
				\multicolumn{2}{r|}{Precision}       & \TextField[name = ps.e.p.chb,width=2.5em]{}   & -  & \TextField[name = ps.e.p.siena,width=2.5em]{}  & \TextField[name = ps.e.p.sz,width=2.5em]{}  & \TextField[name = ps.s.p.chb,width=2.5em]{}   & -  & \TextField[name = ps.s.p.siena,width=2.5em]{}  & \TextField[name = ps.s.p.sz,width=2.5em]{}  \\
				\multicolumn{2}{r|}{FP/day}         & \TextField[name = ps.e.fp.chb,width=2.5em]{}  & -  & \TextField[name = ps.e.fp.siena,width=2.5em]{} & \TextField[name = ps.e.fp.sz,width=2.5em]{} & \TextField[name = ps.s.fp.chb,width=2.5em]{}  & -  & \TextField[name = ps.s.fp.siena,width=2.5em]{} & \TextField[name = ps.s.fp.sz,width=2.5em]{} \\
				\midrule
				\midrule
				\multicolumn{10}{c}{\textit{Performance of a subject-independent model cross-validated on a single dataset}} \\
				\midrule
				\multicolumn{2}{r|}{Metrics} & \multicolumn{4}{c|}{Event-based} & \multicolumn{4}{c}{Sample-based} \\
				\multicolumn{2}{l|}{} & CHB-MIT & TUH & Siena & SeizeIT & CHB-MIT  & TUH & Siena & SeizeIT \\
				\midrule
				\multicolumn{2}{r|}{F1-score}    & \TextField[name = pi.e.f1.chb,width=2.5em]{}  & \TextField[name = pi.e.f1.tuh,width=2.5em]{} & \TextField[name = pi.e.f1.siena,width=2.5em]{} & \TextField[name = pi.e.f1.sz,width=2.5em]{} & \TextField[name = pi.s.f1.chb,width=2.5em]{}  & \TextField[name = pi.s.f1.tuh,width=2.5em]{}  & \TextField[name = pi.s.f1.siena,width=2.5em]{} & \TextField[name = pi.s.f1.sz,width=2.5em]{} \\
				\multicolumn{2}{r|}{Sensitivity} & \TextField[name = pi.e.s.chb,width=2.5em]{}   & \TextField[name = pi.e.s.tuh,width=2.5em]{}  & \TextField[name = pi.e.s.siena,width=2.5em]{}  & \TextField[name = pi.e.s.sz,width=2.5em]{}  & \TextField[name = pi.s.s.chb,width=2.5em]{}   & \TextField[name = pi.s.s.tuh,width=2.5em]{}  & \TextField[name = pi.s.s.siena,width=2.5em]{}  & \TextField[name = pi.s.s.sz,width=2.5em]{}  \\
				\multicolumn{2}{r|}{Precision}   & \TextField[name = pi.e.p.chb,width=2.5em]{}   & \TextField[name = pi.e.p.tuh,width=2.5em]{}  & \TextField[name = pi.e.p.siena,width=2.5em]{}  & \TextField[name = pi.e.p.sz,width=2.5em]{}  & \TextField[name = pi.s.p.chb,width=2.5em]{}   & \TextField[name = pi.s.p.tuh,width=2.5em]{}  & \TextField[name = pi.s.p.siena,width=2.5em]{}  & \TextField[name = pi.s.p.sz,width=2.5em]{}  \\
				\multicolumn{2}{r|}{FP/day}     & \TextField[name = pi.e.fp.chb,width=2.5em]{}  & \TextField[name = pi.e.fp.tuh,width=2.5em]{} & \TextField[name = pi.e.fp.siena,width=2.5em]{} & \TextField[name = pi.e.fp.sz,width=2.5em]{} & \TextField[name = pi.s.fp.chb,width=2.5em]{}  & \TextField[name = pi.s.fp.tuh,width=2.5em]{}  & \TextField[name = pi.s.fp.siena,width=2.5em]{} & \TextField[name = pi.s.fp.sz,width=2.5em]{} \\
				\midrule
				\midrule
				\multicolumn{10}{c}{\textit{Performance of a subject-independent model trained on a independent dataset}} \\
				\midrule
				Training  & Metrics & \multicolumn{4}{c|}{Event-based} & \multicolumn{4}{c}{Sample-based} \\
				Data                              &             & CHB-MIT                                    & TUH                                        & Siena                                        & SeizeIT                                   & CHB-MIT                                    & TUH                                        & Siena                                        & SeizeIT                                   \\
				\midrule
				\TextField[name = di,width=2.5em]{} & F1-score    & \TextField[name = di.e.f1.chb,width=2.5em]{} & \TextField[name = di.e.f1.tuh,width=2.5em]{} & \TextField[name = di.e.f1.siena,width=2.5em]{} & \TextField[name = di.e.f1.sz,width=2.5em]{} & \TextField[name = di.s.f1.chb,width=2.5em]{} & \TextField[name = di.s.f1.tuh,width=2.5em]{} & \TextField[name = di.s.f1.siena,width=2.5em]{} & \TextField[name = di.s.f1.sz,width=2.5em]{} \\
				                                  & Sensitivity & \TextField[name = di.e.s.chb,width=2.5em]{}  & \TextField[name = di.e.s.tuh,width=2.5em]{}  & \TextField[name = di.e.s.siena,width=2.5em]{}  & \TextField[name = di.e.s.sz,width=2.5em]{}  & \TextField[name = di.s.s.chb,width=2.5em]{}  & \TextField[name = di.s.s.tuh,width=2.5em]{}  & \TextField[name = di.s.s.siena,width=2.5em]{}  & \TextField[name = di.s.s.sz,width=2.5em]{}  \\
				                                  & Precision   & \TextField[name = di.e.p.chb,width=2.5em]{}  & \TextField[name = di.e.p.tuh,width=2.5em]{}  & \TextField[name = di.e.p.siena,width=2.5em]{}  & \TextField[name = di.e.p.sz,width=2.5em]{}  & \TextField[name = di.s.p.chb,width=2.5em]{}  & \TextField[name = di.s.p.tuh,width=2.5em]{}  & \TextField[name = di.s.p.siena,width=2.5em]{}  & \TextField[name = di.s.p.sz,width=2.5em]{}  \\
				                                  & FP/day     & \TextField[name = di.e.fp.chb,width=2.5em]{} & \TextField[name = di.e.fp.tuh,width=2.5em]{} & \TextField[name = di.e.fp.siena,width=2.5em]{} & \TextField[name = di.e.fp.sz,width=2.5em]{} & \TextField[name = di.s.fp.chb,width=2.5em]{} & \TextField[name = di.s.fp.tuh,width=2.5em]{} & \TextField[name = di.s.fp.siena,width=2.5em]{} & \TextField[name = di.s.fp.sz,width=2.5em]{} \\
			\end{tabular}
		\end{table}
\end{tcolorbox}
\end{Form}
\end{table}

\begin{table}
\caption{SzCORE reproducibility checklist}
\label{tbl:checklist}
\begin{tcolorbox}[colback=white,colframe=black]
\subsection*{SzCORE Reproducibility Checklist}

For all \textbf{models} and \textbf{algorithms} presented, check if you include:
\begin{todolist}
    \item A clear description of the mathematical setting, algorithm, and/or model including assumptions and parameters.
    \item A description of the input data of the algorithm specifying sampling frequency, and number of channels.
    \item An analysis of the complexity (time, space, sample size) of any algorithm.
\end{todolist}
\bigbreak

For all \textbf{datasets} uses, check if you include:
\begin{todolist}
    \item A description of the dataset including the number of subjects, number of seizures, seizure type, and recording duration.
    \item The details of the train / validation / test splits that respect subject independence and chronology.
    \item An explanation of any data that were excluded, and all pre-processing steps.
    \item A link to a downloadable version of the dataset.
    \item For new data collected, a complete description of the data collection process, such as instructions to annotators and methods for quality control along with a BIDS-EEG / HED-SCORE compatible version of the dataset.
\end{todolist}
\bigbreak

For all share \textbf{code} related to this work, check if you include:
\begin{todolist}
    \item Specification of dependencies.
    \item Training code.
    \item Evaluation code.
    \item (Pre-)trained model(s).
    \item \textsc{README} file includes table of results accompanied by precise command to run to produce results.
\end{todolist}
\bigbreak

For all reported \textbf{experimental results}, check if you include:
\begin{todolist}
    \item The range of hyper-parameters considered, method to select the best hyper-parameter configuration, and specification of all hyper-parameters used to generate results.
    \item The exact number of training and evaluation runs.
    \item A clear definition of the specific measure or statistics used to report results.
    \item A description of results with a report of sensitivity, precision, F1-score and false alarm rate per day.
    \item A description of results on the publicly available datasets, namely Physionet CHB-MIT Scalp EEG Database, TUH EEG Seizure Corpus, Physionet Siena Scalp EEG, and SeizIT1.
    \item The average runtime for each result, or estimated energy cost.
    \item A description of the computing infrastructure used.
\end{todolist}
\end{tcolorbox}
\end{table}

\section{Ways to contribute}\label{A:sec:contribute}

This framework and benchmark should foster reproducible, transparent, and efficient research. It will benefit from contributions from the community. Here, we list several ways to contribute to the framework and benchmark. 
\subsubsection{Feedback}
\begin{itemize}
\item Provide feedback on the proposed framework and benchmark using the following form: \url{https://forms.gle/XfbDaJQi2VooWRN2A}. We are interested in problems and challenges you might have encountered and ideas to improve the framework and the platform. 
\item Get in touch with us if you need help or are interested in a collaboration. You can contact the corresponding author (\href{mailto:jonathan.dan@epfl.ch}{jonathan.dan@epfl.ch}) or any author of the paper.
\end{itemize}

\subsubsection{Contribute to the 10-20 seizure detection benchmark}
\begin{itemize}
\item Report results of your seizure detection and include results of your algorithm on the 10-20 seizure detection benchmark. The checklist and model card in this paper should help you report results in a compliant manner.
\item Submit the performance of your algorithm on the online platform: \url{https://eslweb.epfl.ch/epilepsybenchmarks}. Once validated, your results will be displayed publicly and compared to other algorithms. 
\item Contribute to a new dataset. Our field would benefit from more datasets that can be used for the validation of seizure detection algorithms. We expect high quality scalp-EEG datasets that adhere to the SzCORE framework. These could either be made publicly available or could be used as a private test dataset on the online platform.
\item Contribute to the code libraries that enable the SzCORE framework and benchmark. The different code libraries are open-source and open to community contributions. You will find them on Github: \url{https://github.com/esl-epfl/sz-validation-framework}.
\end{itemize}

\end{document}